\documentclass[10pt, conference]{IEEEtran}
\IEEEoverridecommandlockouts
\usepackage[english]{babel}
\usepackage{amsmath,amssymb,amsfonts,latexsym}
\usepackage[amsthm]{ntheorem}
\usepackage{thmtools}
\declaretheoremstyle[headfont=\scshape\bfseries,qed=\qedsymbol ]{mystyle}
\renewcommand\qedsymbol{$\blacksquare$}

\usepackage{tabularx}

\usepackage[mathscr]{euscript}
\usepackage{graphicx}
\usepackage{gensymb}
\usepackage{balance}
\usepackage[inline]{enumitem}
\usepackage[font=small,skip=0pt,figurename=Fig.]{caption}
\usepackage{microtype}
\usepackage{textcomp}
\usepackage{mathtools, nccmath}
\usepackage{commath}
\usepackage{xcolor}
\usepackage{lettrine}
\usepackage{siunitx}
\usepackage{multirow}
\usepackage{array}
\usepackage{nicefrac}
\usepackage{cite}
\usepackage{verbatim}
\usepackage{hyperref}
\usepackage{lipsum}% Just for this example
\usepackage{graphicx}
\usepackage{caption}
\usepackage{subcaption}
\usepackage{amsmath,amssymb}
\usepackage{amsmath}
\usepackage{amssymb}
\usepackage{mathtools}
\usepackage[utf8]{inputenc}
\usepackage{amsmath}
\usepackage{algorithm}
\usepackage{algpseudocode}
\makeatletter
\algrenewcommand\alglinenumber[1]{\footnotesize\arabic{ALG@line}:}
\makeatother
\usepackage{algorithm}
\usepackage{algpseudocode}
\newcolumntype{M}[1]{>{\centering\arraybackslash}m{#1}}

\allowdisplaybreaks

\begin{document}

\title{Deep Unfolding for SIM-Assisted Multiband MU-MISO Downlink Systems}

\author{\IEEEauthorblockN{
Muhammad Ibrahim, Amine Mezghani, and Ekram Hossain, {\em FIEEE}\thanks{The authors are with the Department of Electrical and Computer Engineering at the University of Manitoba, MB, Canada (emails: ibrah101@myumanitoba.ca, \{Amine.Mezghani, Ekram.Hossain\}@umanitoba.ca).}
}}

\maketitle

\begin{abstract}
To improve the efficiency of scarce radio-frequency (RF) resources in next-generation wireless systems, an intelligent transceiver architecture based on stacked intelligent metasurfaces (SIM) has recently emerged, where multiple programmable metasurface layers are cascaded and each layer consists of passive meta-atoms that perform beamforming directly in the wave domain. In parallel, inter-band carrier aggregation enables multi-band transmission with high spectral efficiency. Their integration in multi-band multiuser downlink transmission is challenging because a single SIM phase configuration must remain effective across all subcarriers, while user scheduling and power allocation vary across scheduling intervals. To address these challenges, we propose an alternating-optimization framework that decomposes the joint design into a power-constrained precoder update and a SIM phase update. For the SIM phase subproblem, we develop a physically consistent multi-band deep-unfolding network (MBDU-Net) that unrolls projected-gradient phase updates into a compact trainable architecture. Each stage computes an analytic gradient from the cascaded SIM channel model and learns lightweight parameters, including per-stage step sizes and band-aware scaling, enabling fast convergence. Numerical results for multi-band multiuser downlink scenarios demonstrate convergence and consistent sum-rate gains on unseen channel realizations.
\end{abstract}

\vspace{-12pt}

\section{Introduction}

Programmable metasurfaces enable dynamic control of electromagnetic (EM) waves by electronically tuning the responses of densely packed sub-wavelength meta-atoms distributed over a planar aperture \cite{di2020smart}. This capability has stimulated substantial interest in metasurface-assisted wireless communications, where the propagation channel and the transceiver front-end can be reconfigured to improve spectral efficiency with low power consumption \cite{ibrahim2024performance}. More recently, a new concept based on SIM has been introduced, in which multiple transmissive reconfigurable intelligent surface (RIS) layers are cascaded and placed in close proximity to the base station (BS) \cite{an2024stacked}. By embedding passive, phase-programmable layers into the BS architecture, SIM implements wave-domain beamforming and can reduce the reliance on high-dimensional digital precoding and many RF chains. In this architecture, each layer applies an adjustable phase shift to the impinging wavefront, providing controllable end-to-end beamforming for multiuser transmission. {Recent studies have explored SIM-assisted transceiver design \cite{an2024stacked}, holographic MIMO communications \cite{papazafeiropoulos2024achievable}, physically consistent multiport modeling \cite{abrardo2025novel}, and wideband multi-user transmission \cite{ibrahim2026stacked}.}

In parallel, multi-band carrier aggregation (CA) has become a key mechanism for improving spectral efficiency by jointly exploiting subcarriers across multiple frequency bands \cite{khan2022study}. Compared with single-band operation, multi-band CA can better leverage frequency-selective fading and band diversity, where lower bands typically offer more favorable propagation and reliability, while higher bands provide wider bandwidth but may suffer from higher noise power and less robust scattering \cite{balasuriya2025physically}. In downlink multiuser systems, these heterogeneous band characteristics provide additional degrees of freedom for subcarrier selection and power allocation, leading to sum-rate gains.

Integrating SIM-assisted transmission with multi-band CA, however, is challenging because the SIM phase configuration is effectively shared across the aggregated spectrum, whereas the per-subcarrier operating conditions vary substantially across bands. In particular, band-dependent bandwidths and noise levels, together with subcarrier-dependent interference shaped by multiuser scheduling and power allocation, induce gradients of markedly different scales across the low- and high-band components. Consequently, SIM phase design in a multi-band multiuser downlink leads to a tightly coupled optimization problem that is non-convex and strongly frequency coupled.
Classical optimization methods for SIM-assisted multiuser transmission typically adopt alternating optimization, where the digital precoders are updated for a fixed SIM configuration and the SIM phases are then refined for fixed precoders. 
%The SIM update is commonly implemented through projected-gradient iterations under the unit-modulus phase constraints. Although this approach is conceptually transparent and directly tied to the underlying objective, 
However, this approach can be computationally demanding in practical multiuser, multi-subcarrier systems. 

Recently, data-driven methods have been introduced for physical-layer optimization, where a neural network maps channel realizations to transceiver parameters \cite{xia2019deep}, such as beamforming weights or metasurface configurations. Deep unfolding improves interpretability by unrolling iterative optimization algorithms into trainable architectures \cite{pellaco2021matrix}. {However, most unfolding-based beamforming and phase-optimization methods are developed for conventional MIMO/RIS systems and do not explicitly address SIM-assisted multi-band transmission.}

Motivated by these considerations, we propose an alternating-optimization framework for SIM-assisted multi-band MU-MISO downlink transmission, where we optimize digital precoding and SIM phase shifts. For the SIM phase optimization subproblem, we develop a physically-consistent multi-band deep-unfolding network (MBDU-Net). The key idea is to unroll a fixed number of projected-gradient phase updates into a finite-depth architecture, where each stage preserves the underlying update structure while replacing hand-tuned update parameters with lightweight trainable variables. {Unlike standard unfolding designs with a single learned update rule, MBDU-Net separates the low- and high-band gradient contributions and learns band-aware step sizes together with momentum-related parameters.} Training is performed offline using a rate-based objective, whereas online inference has fixed complexity and avoids running a large number of iterations.

%-------------------------
\noindent
\textbf{Notations:} For any matrix $\mathbf{A}$, $\mathbf{A}^T$, $\mathbf{A}^*$, and $\mathbf{A}^H$ represent its transpose, conjugate, and conjugate transpose, respectively. $\mathrm{Diag}(\mathbf{a})$ denotes a diagonal matrix with the elements of vector $\mathbf{a}$ along its diagonal, while $\mathrm{diag}(\mathbf{A})$ returns a vector containing the diagonal elements of the matrix $\mathbf{A}$.

%--------------------------------------------------------------------------

\section{System Model and Assumptions}
\begin{figure}[tb!]
\centerline{\includegraphics[width = 9cm]{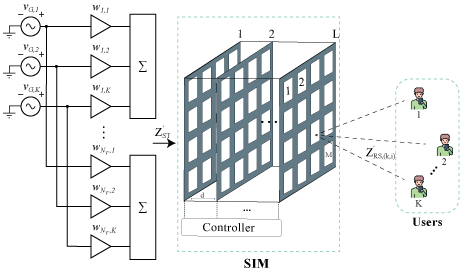}}
\vspace{4pt}
\caption{SIM-aided multi-user MISO system }
\label{fig:1}
\end{figure}

We consider a SIM-assisted multi-band multiuser downlink (Fig.~\ref{fig:1}), where SIM is integrated into the BS architecture. The BS employs a uniform linear array (ULA) of $N_t$ active antennas and a transmissive SIM comprising $L$ cascaded layers, each with $M$ passive meta-atoms (elements). The system operates over two bands using inter-band carrier aggregation, with $M_L$ and $M_H$ orthogonal subcarriers allocated to the low- and high-frequency bands, respectively. The corresponding per-subcarrier bandwidths are denoted by $B_L$ and $B_H$. The BS operates under a total downlink transmit power budget $P_T$. Owing to the larger bandwidth in the high band, we allow different noise variances across bands and model the additive white Gaussian noise (AWGN) power as $\sigma_H^2 > \sigma_L^2$. For notational convenience, we use a unified subcarrier index $i$ across both bands and define the overall frequency set as $f_i \in \mathcal{F}=\{f_{L,1},\ldots,f_{L,M_L},\, f_{H,1},\ldots,f_{H,M_H}\}$.

%--------------------------------------------------------------------------

\subsection{Physically-Consistent SIM Model}
We model SIM as a cascade of $L$ transmissive RIS layers, where each layer contains $M$ passive meta-atoms and is controlled by a phase vector $\boldsymbol{\phi}_{\ell}=[\phi_{\ell,1},\ldots,\phi_{\ell,M}]^{\top}$, $\ell\in\{1,\ldots,L\}$. Following the physically-consistent circuit-theoretic formulation in \cite{ibrahim2026stacked}, the SIM is characterized by two banded impedance matrices: (i) the interconnection impedance matrix $\mathbf{Z}_{SS}(f)$, which captures mutual coupling and propagation between adjacent layers, and (ii) the load impedance matrix $\mathbf{Z}_{S}(\boldsymbol{\Phi})$, which captures the programmable impedances within each layer. Assuming $M$ meta-atoms per layer, each layer is modeled as a $2M$-port network and the overall SIM has $2ML$ ports. The SIM transfer matrix at frequency $f$ is $ \label{eq:G_def_conf}
\mathbf{G}(\boldsymbol{\Phi},f)
=
\Big(\mathbf{Z}_{SS}(f)+\mathbf{Z}_{S}(\boldsymbol{\Phi})\Big)^{-1}$, where $\mathbf{Z}_{SS}(f)$ accounts for the mutual coupling and $\mathbf{Z}_{S}(\boldsymbol{\Phi})$ accounts for the programmable load network. Both $\mathbf{Z}_{SS}(f)$ and $\mathbf{Z}_{S}(\boldsymbol{\Phi})$ exhibit a block-banded structure and can be decomposed into $M\times M$ submatrices~\cite{abrardo2025novel}. {The detailed circuit-theoretic SIM model, including the explicit block-banded form of \(\mathbf{Z}_{SS}(f)\) and its frequency-dependent inter-layer coupling blocks, is provided in \cite{ibrahim2026stacked}.} The programmable load matrix $\mathbf{Z}_{S}(\boldsymbol{\Phi})$ is:
%In particular, $\mathbf{Z}_{SS}(f)$ can be written as
%\begin{equation}
%\label{eq:ZSS_block_conf}
%\mathbf{Z}_{SS}(f)=
%\begin{bmatrix}
%\mathbf{Z}^{(0)}_{2,2}(f) & \mathbf{0} & \mathbf{0} & \cdots & \mathbf{0}\\
%\mathbf{0} & \mathbf{Z}^{(1)}_{1,1}(f) & \mathbf{Z}^{(1)}_{1,2}(f) & \cdots & \mathbf{0}\\
%\mathbf{0} & \mathbf{Z}^{(1)}_{2,1}(f) & \mathbf{Z}^{(1)}_{2,2}(f) & \cdots & \mathbf{0}\\
%\vdots & \vdots & \vdots & \ddots & \vdots\\
%\mathbf{0} & \mathbf{0} & \mathbf{0} & \cdots & \mathbf{Z}^{(L)}_{1,1}(f)
%\end{bmatrix},
%\end{equation}
%where $\mathbf{Z}^{(\ell)}_{i,j}(f)\in\mathbb{C}^{M\times M}$, $i,j\in\{1,2\}$, describes the coupling between consecutive layers, while $\mathbf{Z}^{(0)}_{2,2}(f)$ and $\mathbf{Z}^{(L)}_{1,1}(f)$ represent the terminal port blocks of the first and last layers, respectively. 

\begin{equation}
\label{eq:ZS_block_conf}
\mathbf{Z}_{S}(\boldsymbol{\Phi})=
\begin{bmatrix}
\mathbf{X}^{(1)}_{1,1} & \mathbf{X}^{(1)}_{1,2} & \mathbf{0} & \cdots & \mathbf{0}\\
\mathbf{X}^{(1)}_{2,1} & \mathbf{X}^{(1)}_{2,2} & \mathbf{0} & \cdots & \mathbf{0}\\
\mathbf{0} & \mathbf{0} & \mathbf{X}^{(2)}_{1,1} & \ddots & \vdots\\
\vdots & \vdots & \ddots & \ddots & \mathbf{0}\\
\mathbf{0} & \mathbf{0} & \cdots & \mathbf{X}^{(L)}_{2,1} & \mathbf{X}^{(L)}_{2,2}
\end{bmatrix},
\end{equation}
where, for notational simplicity, we omit the explicit dependence $\mathbf{X}^{(\ell)}_{i,j}=\mathbf{X}^{(\ell)}_{i,j}(\boldsymbol{\phi}_\ell)$. We assume phase-only control, so each load submatrix $\mathbf{X}^{(\ell)}_{i,j}$ is diagonal, with diagonal entries $x^{(\ell)}_{m}(i,j)$ determined by the per-element phase shifts, $m=1,\ldots,M$, and $i,j\in\{1,2\}$.

%\label{eq:X_diag_conf} Using the standard two-port model of a transmissive meta-atom, the corresponding impedance matrix for the $m$-th element in layer $\ell$ is
%\begin{equation}
%\label{eq:Zm_conf}
%\mathbf{Z}^{(\ell)}_{m}
%=
%j Z_{0}
%\begin{bmatrix}
%\dfrac{\cos(\phi_{\ell,m})}{\sin(\phi_{\ell,m})} & \dfrac{1}{\sin(\phi_{\ell,m})}\\[6pt]
%\dfrac{1}{\sin(\phi_{\ell,m})} & \dfrac{\cos(\phi_{\ell,m})}{\sin(\phi_{\ell,m})}
%\end{bmatrix},
%\end{equation}
%and each diagonal block in \eqref{eq:X_diag_conf} is obtained by collecting the corresponding entries of $\mathbf{Z}^{(\ell)}_{m}$ for $m=1,\ldots,M$.

%--------------------------------------------------------------------------

\subsection{Signal Transmission Model}
\label{Sub:2B}
Consider subcarrier $i$ with center frequency $f_i\in\mathcal{F}$. The BS transmits $K$ independent data streams collected in
$\mathbf{s}_i=[s_{i,1},\ldots,s_{i,K}]^{\top}\in\mathbb{C}^{K\times 1}$, where $\mathbb{E}\{\mathbf{s}_i\mathbf{s}_i^{H}\}=\mathbf{I}_K$.
Let $\mathbf{W}_i=[\mathbf{w}_{i,1},\ldots,\mathbf{w}_{i,K}]\in\mathbb{C}^{N_t\times K}$ denote the linear precoder on subcarrier $i$.
The transmitted signal is $\label{eq:xi_def}
\mathbf{x}_i=\mathbf{W}_i\mathbf{s}_i\in\mathbb{C}^{N_t\times 1}$. {Thus, the model uses one independent stream per scheduled user on each subcarrier, and users may receive independent symbols over subcarriers assigned across both bands.}

Following \cite{gradoni2021end}, the end-to-end transfer function from the BS ports to the user ports can be expressed in the impedance domain as $\label{eq:HZ_def}
\mathbf{H}_{Z,i}(\boldsymbol{\Phi})
=
\mathbf{Z}_{RS,i}\,\mathbf{G}(\boldsymbol{\Phi},f_i)\,\mathbf{Z}_{ST,i}$, where $\mathbf{Z}_{ST,i}$ denotes the impedance coupling matrix from the BS antenna ports to the SIM, and $\mathbf{Z}_{RS,i}$ denotes the impedance coupling matrix from the SIM to the user ports, on subcarrier $i$. Since only the first face of the first SIM layer is excited by the BS and only the last face of the last SIM layer is connected to the receivers, the effect of SIM can be written in terms of the submatrix $\mathbf{G}_{2L,1}(\boldsymbol{\Phi},f_i)\in\mathbb{C}^{M\times M}$ as
\begin{equation}
\label{eq:Heq_Zform}
\mathbf{H}_i(\boldsymbol{\Phi})
=
\mathbf{Z}'_{RS,i}\,\mathbf{G}_{2L,1}(\boldsymbol{\Phi},f_i)\,\mathbf{Z}'_{ST,i},
\end{equation}
where $\mathbf{Z}'_{ST,i}\in\mathbb{C}^{M\times N_t}$ is formed by the first $M$ rows of $\mathbf{Z}_{ST,i}$ corresponding to the excited ports at the first SIM face, and $\mathbf{Z}'_{RS,i}\in\mathbb{C}^{K\times M}$ is formed by the last $M$ columns of $\mathbf{Z}_{RS,i}$ corresponding to the observed ports at the last SIM face. Here, $\mathbf{G}_{2L,1}(\boldsymbol{\Phi},f_i)$ is an impedance-domain transfer matrix mapping the port quantities at the first SIM face to those at the last SIM face through the stacked structure. Accordingly, the received signal on subcarrier $i$ is $
\label{eq:yi_def}
\mathbf{y}_i
=
\mathbf{H}_i(\boldsymbol{\Phi})\,\mathbf{x}_i+\mathbf{n}_i
= \mathbf{H}_i(\boldsymbol{\Phi})\,\mathbf{W}_i\mathbf{s}_i+\mathbf{n}_i$ where $\mathbf{y}_i\in\mathbb{C}^{K\times 1}$ and $\mathbf{n}_i\sim\mathcal{CN}(\mathbf{0},\sigma_i^2\mathbf{I}_K)$ denotes AWGN, with $\sigma_i^2\in\{\sigma_L^2,\sigma_H^2\}$ depending on whether $f_i$ lies in the low or high band. Inter-user interference is treated as colored noise. Accordingly, the interference-plus-noise covariance matrix in the transmit domain for user $k$ on subcarrier $i$ is
\begin{equation}
\label{eq:Cik_def}
\mathbf{C}_{i,k}
=
\sigma_i^{2}\mathbf{I}_{N_t}
+\sum_{\substack{j=1\\ j\neq k}}^{K}
\mathbf{h}_{i,j}^{H}\mathbf{w}_{i,j}\mathbf{w}_{i,j}^{H}\mathbf{h}_{i,j},
\qquad
\mathbf{C}_{i,k}\in\mathbb{C}^{N_t\times N_t},
\end{equation}
where $\mathbf{h}_{i,k}\in\mathbb{C}^{1\times N_t}$ denotes the effective channel of user $k$ on subcarrier $i$ (i.e., the $k$-th row of $\mathbf{H}_i(\boldsymbol{\Phi})$) and $\mathbf{w}_{i,k}\in\mathbb{C}^{N_t\times 1}$ denotes its beamformer. 
%The achievable sum rate of the SIM-assisted multi-band downlink can then be expressed in determinant form as
%\begin{align}
%\label{eq:R_det_form}
%R
%&=\sum_{i=1}^{M_L+M_H}\sum_{k=1}^{K} B_i
%\Big[
%\log_{2}\!\big|\mathbf{C}_{i,k}
%+ %\mathbf{h}_{i,k}^{H}\mathbf{w}_{i,k}\mathbf{w}_{i,k}^{H}\mathbf{h}_{i,k}\big|
%\nonumber\\
%&\hspace{26mm}
%-\log_{2}\!\big|\mathbf{C}_{i,k}\big|
%\Big].
%\end{align}
Since $\mathbf{C}_{i,k}$ is positive semi-definite, using the eigendecomposition
$\mathbf{C}_{i,k}=\mathbf{U}_{i,k}\boldsymbol{\Lambda}_{i,k}\mathbf{U}_{i,k}^{H}$, we define the whitened channel
$\widetilde{\mathbf{h}}_{i,k}
=\mathbf{h}_{i,k}\,\mathbf{U}_{i,k}\boldsymbol{\Lambda}_{i,k}^{-1/2}$.
Applying the matrix determinant lemma, the sum rate can be rewritten as
\begin{align}
\label{eq:R_scalar_form}
R
&=\sum_{i=1}^{M_L+M_H}\sum_{k=1}^{K} B_i\,
\log_{2}\!\big|
 \mathbf{I}+ \widetilde{\mathbf{h}}_{i,k}^{H}\,\mathbf{w}_{i,k}\mathbf{w}_{i,k}^{H}\,\widetilde{\mathbf{h}}_{i,k}\big|.
\end{align}

Finally, we adopt a Rayleigh fading model for the transimpedance coupling matrices. In particular, for $\mathbf{Z}'_{RS,i}$ we use the multiport model in \cite{ivrlavc2014multiport}:
\begin{equation}
\label{eq:ZRS_rayleigh}
\mathbf{Z}'_{RS,i}
=
\frac{c}{2\pi f_i\, d^{\gamma/2}}
\big(\Re\{\mathbf{Z}_{R}(f_i)\}\big)^{1/2}\,
\mathbf{F}_i\,
\big(\Re\{\mathbf{Z}_{S}(f_i)\}\big)^{1/2},
\end{equation}
where $c$ is the speed of light, $d$ is the link distance, $\gamma$ is the path-loss exponent, and $\mathbf{F}_i\in\mathbb{C}^{K\times M}$ has i.i.d.\ $\mathcal{CN}(0,1)$ entries. Assuming negligible mutual coupling at the user side and frequencies in the GHz range, $\big(\Re\{\mathbf{Z}_{R}(f_i)\}\big)^{1/2}$ can be approximated by $\sqrt{R_a}\,\mathbf{I}$, where $R_a$ denotes the antenna resistance. Additonally, in this paper, $\mathbf{Z}_{ST}$ and the related submatrices %$\mathbf{Z}_{i,j}^{(\ell)}$ in \eqref{eq:ZSS_block_conf} 
are computed using the method proposed in~\cite{akrout2023super}.

%--------------------------------------------------------------------------

\section{Optimization Problem and Solution}
\noindent\textit{A) Problem Formulation:}
Given the multi-band signal model in Section~II.B, our goal is to jointly design the per-subcarrier precoders $\{\mathbf{W}_i\}_{i=1}^{M_L+M_H}$ and the SIM phase configuration $\boldsymbol{\Phi}$ to maximize the multi-band sum rate. In particular, using the 
%determinant-form 
rate expression in 
%\eqref{eq:R_det_form} (equivalently 
\eqref{eq:R_scalar_form}, we formulate
\begin{equation}
\label{eq:opt_main}
\begin{aligned}
\max_{\{\mathbf{W}_i\},\,\boldsymbol{\Phi}}\quad 
& R(\{\mathbf{W}_i\},\boldsymbol{\Phi}) \\
\text{s.t.}\quad 
& \sum_{i=1}^{M_L+M_H}\operatorname{tr}\!\big(\mathbf{W}_i\mathbf{W}_i^{H}\big)\le P_T,\\
& 0\le \phi_{\ell,m}<2\pi,\qquad \forall\,\ell,\ m .
\end{aligned}
\end{equation}

%----------------------------------------
\noindent\textit{B) Proposed Solution:}
The joint design problem in \eqref{eq:opt_main} is non-convex because the multiuser precoders $\{\mathbf{W}_i\}$ and the SIM phases $\boldsymbol{\Phi}$ are coupled through the effective channels. To handle this coupling, we employ an alternating-optimization procedure that decomposes the problem into two subproblems. In \textbf{P1}, for fixed $\boldsymbol{\Phi}$, the precoders $\{\mathbf{W}_i\}$ are optimized under the total transmit-power constraint. In \textbf{P2}, for fixed precoders, the SIM phase matrix $\boldsymbol{\Phi}$ is updated under the phase-only constraints $0\leq \phi_{\ell,m}<2\pi$. The two subproblems are then solved alternately until convergence.

%--------------------------------------------------------------------------
\noindent\hspace{1em}\textit{1) Power Allocation and Beamforming Optimization:}
For fixed SIM phases $\boldsymbol{\phi}$, subproblem \textbf{P1} optimizes the precoders under the total power constraint. Treating multiuser interference as colored noise with covariance $\mathbf{C}_{i,k}$ in \eqref{eq:Cik_def}, we define the effective gain as $\lambda_{k,i}
=\|\widetilde{\mathbf h}_{i,k}\|^{2}
=\mathbf h_{i,k}\mathbf C_{i,k}^{-1}\mathbf h_{i,k}^{H}$.
The resulting power-allocation problem is a standard water-filling problem over the effective user-subcarrier gains \(\lambda_{k,i}\). Specifically, the powers \(\{p_{k,i}\}\) are selected to maximize the weighted sum rate \(\sum_{i=1}^{N_f}\sum_{k=1}^{K} B_i\log_2(1+\lambda_{k,i}p_{k,i})\), subject to the total power constraint \(\sum_{i=1}^{N_f}\sum_{k=1}^{K}p_{k,i}\le P_{\mathrm{tot}}\) and \(p_{k,i}\ge0\). This problem is solved by iterative water-filling as
\begin{equation}
\label{power}
p_{k,i}^{(t+1)}=\Big[\mu^{(t)}-\frac{1}{\lambda_{k,i}^{(t)}}\Big]^+,
\end{equation}
where $\mu^{(t)}$ is chosen such that $\sum_{i,k} p_{k,i}^{(t+1)}=P_{\mathrm{tot}}$. For each $(k,i)$, the beamformer is selected by matched filtering to the whitened channel, i.e.,
\begin{equation}
\label{eq:w_svd_update}
\mathbf w_{k,i}^{(t+1)}
=
\sqrt{p_{k,i}^{(t+1)}}\,
\frac{\widetilde{\mathbf h}_{i,k}^{H}}{\|\widetilde{\mathbf h}_{i,k}\|_2}.
\end{equation}
The updates are repeated until convergence.

%--------------------------------------------------------------------------

\noindent\hspace{1em}\textit{2) Multi-Band Deep Unfolding for Phase Shift Optimization:}
%\paragraph{Classical Optimization}
A standard approach is projected gradient ascent:
\begin{equation}
\label{eq:phi_pgd_classic}
\boldsymbol{\Phi}^{(t+1)}
=
\Pi_{[0,2\pi)}\!\Big(
\boldsymbol{\Phi}^{(t)} + \eta_t\,\nabla_{\boldsymbol{\Phi}} R
\Big),
\end{equation}
where $\eta_t>0$ is the step size and $\Pi_{[0,2\pi)}(\cdot)$ denotes element-wise projection onto $[0,2\pi)$.  We can compute $\nabla_{\boldsymbol{\Phi}}R$ as
\begin{equation}
\label{eq:dR_dphi_conf}
\frac{\partial R}{\partial \phi_{\ell,m}}
=
\sum_{i=1}^{M_L+M_H}\sum_{k=1}^{K}
\frac{B_i}{\ln 2}\,
\frac{1}{1+s_{i,k}}\,
\frac{\partial s_{i,k}}{\partial \phi_{\ell,m}}.
\end{equation}
While \eqref{eq:phi_pgd_classic} is interpretable and leverages physical structure, its convergence speed is highly sensitive to the step-size and typically requires many iterations, especially in the multi-band system where a single $\boldsymbol{\Phi}$ must jointly accommodate heterogeneous propagation and noise levels across $\mathcal{F}$. 

%This motivates the learning-based approach in the next section, where we unroll it into a trainable multi-band deep-unfolding architecture with data-driven parameters while retaining the same physics-based gradient backbone.

%\paragraph{Multi-Band Deep Unfolding}
 To obtain a fixed-complexity design with faster convergence, we unfold it into a $T$-stage network. In the basic deep-unfolding variant, each stage applies the projected update, but replaces the hand-tuned step size by a trainable parameter $\eta_t$, $t=0,\ldots,T-1$. At stage $t$, the gradient $\nabla_{\boldsymbol{\Phi}}R$ is computed by accumulating the per-subcarrier contributions in \eqref{eq:R_scalar_form}, so the unfolded model preserves the original update structure while learning effective step-sizes. A limitation of basic deep unfolding is that a single step size per stage must accommodate heterogeneous subcarriers across $\mathcal{F}$. In particular, the low-band and high-band components can exhibit different gradient magnitudes due to distinct bandwidths and noise levels, which may lead to conservative updates or oscillatory behavior. To better match this multi-band heterogeneity and accelerate convergence without increasing the number of unfolding stages, we adopt a momentum-accelerated unfolding inspired by the Nesterov acceleration methods \cite{nesterov1983method}.

The proposed update introduces a velocity state $\mathbf{V}^{(t)}$, which has the same dimension as $\boldsymbol{\Phi}$ and aggregates past ascent directions. At stage $t$, we first form a look-ahead phase point by extrapolating along the current velocity. The gradients of the low-band and high-band objectives are then evaluated at this look-ahead point and combined using separate trainable step sizes. Next, the velocity is updated through a trainable momentum factor, and the phase matrix is projected onto the feasible interval. The resulting $T$-stage update is given by
\begin{align}
\label{eq:mbdu_look2}
\widetilde{\boldsymbol{\Phi}}^{(t)}
&=
\boldsymbol{\Phi}^{(t)}+\xi_t\,\mathbf{V}^{(t)},\\
\label{eq:mbdu_dir2}
\mathbf{D}^{(t)}
&= \eta_{L,t}\,\nabla_{\boldsymbol{\Phi}} R_L\!\big(\widetilde{\boldsymbol{\Phi}}^{(t)}\big)
+ \eta_{H,t}\,\nabla_{\boldsymbol{\Phi}} R_H\!\big(\widetilde{\boldsymbol{\Phi}}^{(t)}\big),\\
\label{eq:mbdu_mom2}
\mathbf{V}^{(t+1)}
&= \theta_t\,\mathbf{V}^{(t)} + \mathbf{D}^{(t)},\\
\label{eq:mbdu_upd2}
\boldsymbol{\Phi}^{(t+1)}
&= \Pi_{[0,2\pi)}\!\Big(\widetilde{\boldsymbol{\Phi}}^{(t)}+\mathbf{V}^{(t+1)}\Big),
\end{align}
with initialization $\mathbf{V}^{(0)}=\mathbf{0}$. Here, $\Pi_{[0,2\pi)}(\cdot)$ denotes element-wise projection onto $[0,2\pi)$.

In \eqref{eq:mbdu_dir2}, $R_L$ and $R_H$ follow directly from \eqref{eq:R_scalar_form} by restricting the subcarrier summation to the low-band and high-band index sets, respectively. The parameters $\eta_{L,t}$ and $\eta_{H,t}$ control the step size of the two bands and compensate for their different bandwidth and noise conditions, while $\theta_t$ governs the amount of momentum carried across stages. The extrapolation parameter $\xi_t$ determines the look-ahead point used to evaluate the gradients. Collectively, $\xi_t$, $\theta_t$, $\eta_{L,t}$, and $\eta_{H,t}$ increase the degrees of freedom of the unfolded solver, enabling larger progress per stage and earlier convergence for a fixed depth $T$, instead of relying on a large number of projected-gradient iterations. In the overall AO, the above $T$-stage unfolded block is further unrolled over $I_{\max}$ outer iterations, where each outer iteration contains a depth-$T$ phase-update module. Moreover, each outer iteration is assigned its own trainable parameters. Hence, for outer iteration $i$, the corresponding parameter set is $\{\xi_{i,t},\theta_{i,t},\eta_{L,i,t},\eta_{H,i,t}\}_{t=0}^{T-1}$, and the full architecture consists of $I_{\max}$ cascaded unfolded blocks, each having depth $T$.

Let $\boldsymbol{\Phi}^{(I_{\max},T)}$ denote the network output after $I_{\max}$ outer iterations, each containing $T$ unfolded stages. The trainable parameters are learned offline by minimizing the negative sum rate induced by $\boldsymbol{\Phi}^{(I_{\max},T)}$ over a set of channel realizations, that is, $\mathcal{L} = -R\!\big(\boldsymbol{\Phi}^{(I_{\max},T)}\big)$. During training, the parameters $\{\xi_{i,t},\theta_{i,t},\eta_{L,i,t},\eta_{H,i,t}\}_{i=0}^{I_{\max}-1}{}_{,\,t=0}^{T-1}$ are updated using the Adam optimizer, with gradients obtained by backpropagation through all $T$ unfolded stages across the $I_{\max}$ outer AO iterations.
%---------------------
% Preamble:
% In document:
\begin{algorithm}[t]
\caption{Overall Optimization Algorithm}
\label{alg:ao_mbdu}
\begin{algorithmic}[1]
\Require Total power $P_T$, outer iterations $I_{max}$, unfolding depth $T$
\State Initialize $\boldsymbol{\Phi}$ and $\{\mathbf{W}_i\}$
\For{$i=1$ to $I_{max}$}

  %\State \textbf{P1: Power allocation and beamforming (fixed $\boldsymbol{\Phi}$)}
  \State Compute \{$\mathbf{H}_i(\boldsymbol{\Phi})$\} (\autoref{Sub:2B})
  \State Allocate power $\{p_{k,i}\}$ using Eq. \eqref{power}
  \State Update beamformers $\{\mathbf{w}_{i,k}\}$ using \eqref{eq:w_svd_update}
  %\State \textbf{P2: SIM phase update via unfolding (fixed $\{\mathbf{W}_i\}$)}
  \State Set $\mathbf{V}^{(0)}=\mathbf{0}$ and $\boldsymbol{\Phi}^{(0)}=\boldsymbol{\Phi}$
  \For{$t=0$ to $T-1$}
    \State Form look-ahead point $\widetilde{\boldsymbol{\Phi}}^{(t)}$  (Eq. \eqref{eq:mbdu_look2})
    \State Compute $\nabla_{\boldsymbol{\Phi}} R_L$ and $\nabla_{\boldsymbol{\Phi}} R_H$ (Eq. \eqref{eq:dR_dphi_conf})
    \State Compute $\mathbf{D}^{(t)}$ (Eq. \eqref{eq:mbdu_dir2})
    \State Update $\mathbf{V}^{(t+1)}$ (Eq. \eqref{eq:mbdu_mom2})
    \State Update $\boldsymbol{\Phi}^{(t+1)}$ (Eq. \eqref{eq:mbdu_upd2})
  \EndFor
  \State Set $\boldsymbol{\Phi}\leftarrow \boldsymbol{\Phi}^{(T)}$

\EndFor
\end{algorithmic}
\end{algorithm}
%--------------------------------------------------------------------------
\section{Simulation Results and Discussion}
\begin{figure}[t]
    \centering
    \begin{subfigure}[t]{0.48\linewidth}
        \centering
        \includegraphics[width=\linewidth]{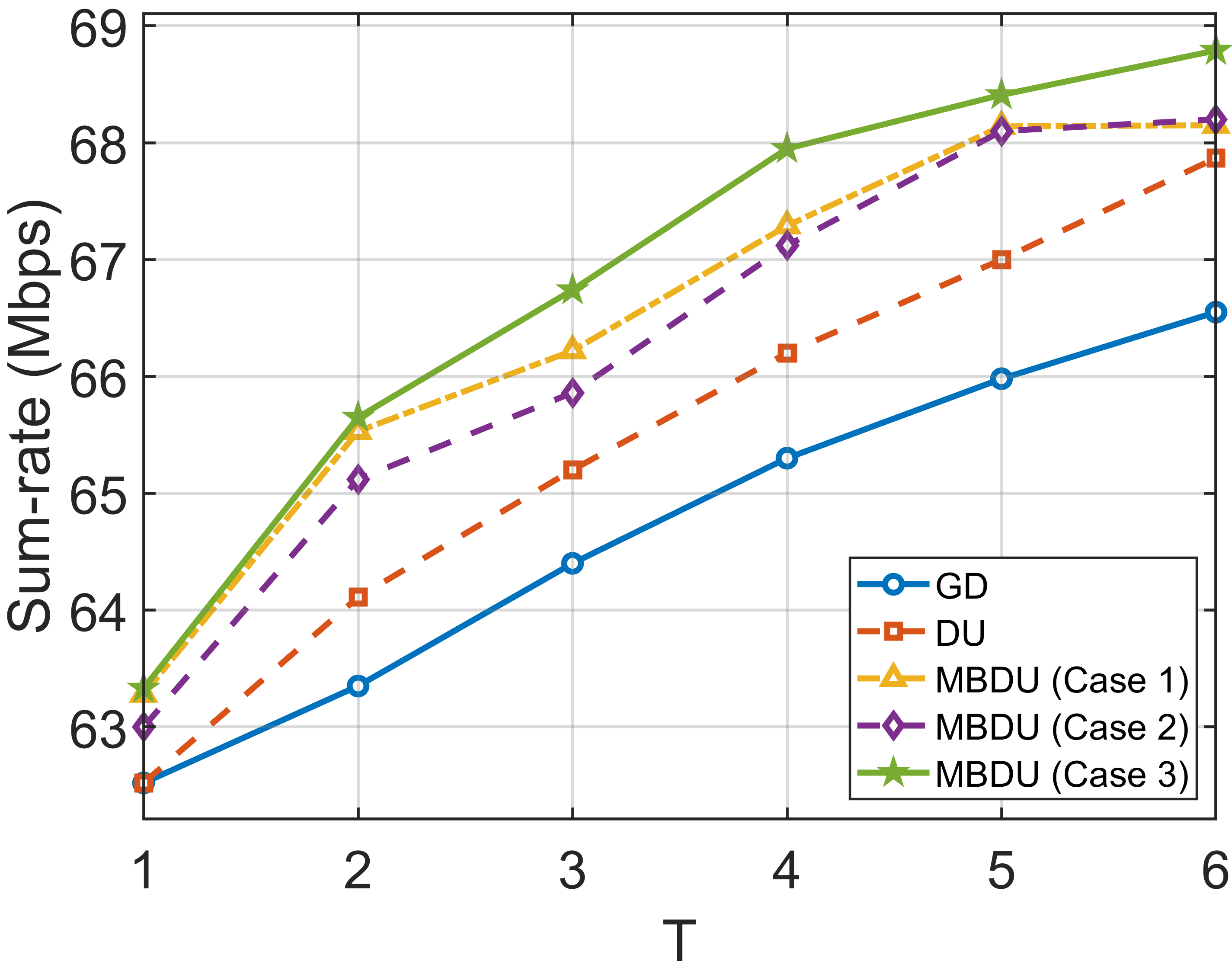}
        \phantomsubcaption\label{fig:T_vs_sumrate_I1}
        \vspace{-1pt}
        {\centering\small (a)\par}
    \end{subfigure}
    \hfill
    \begin{subfigure}[t]{0.48\linewidth}
        \centering
        \includegraphics[width=\linewidth]{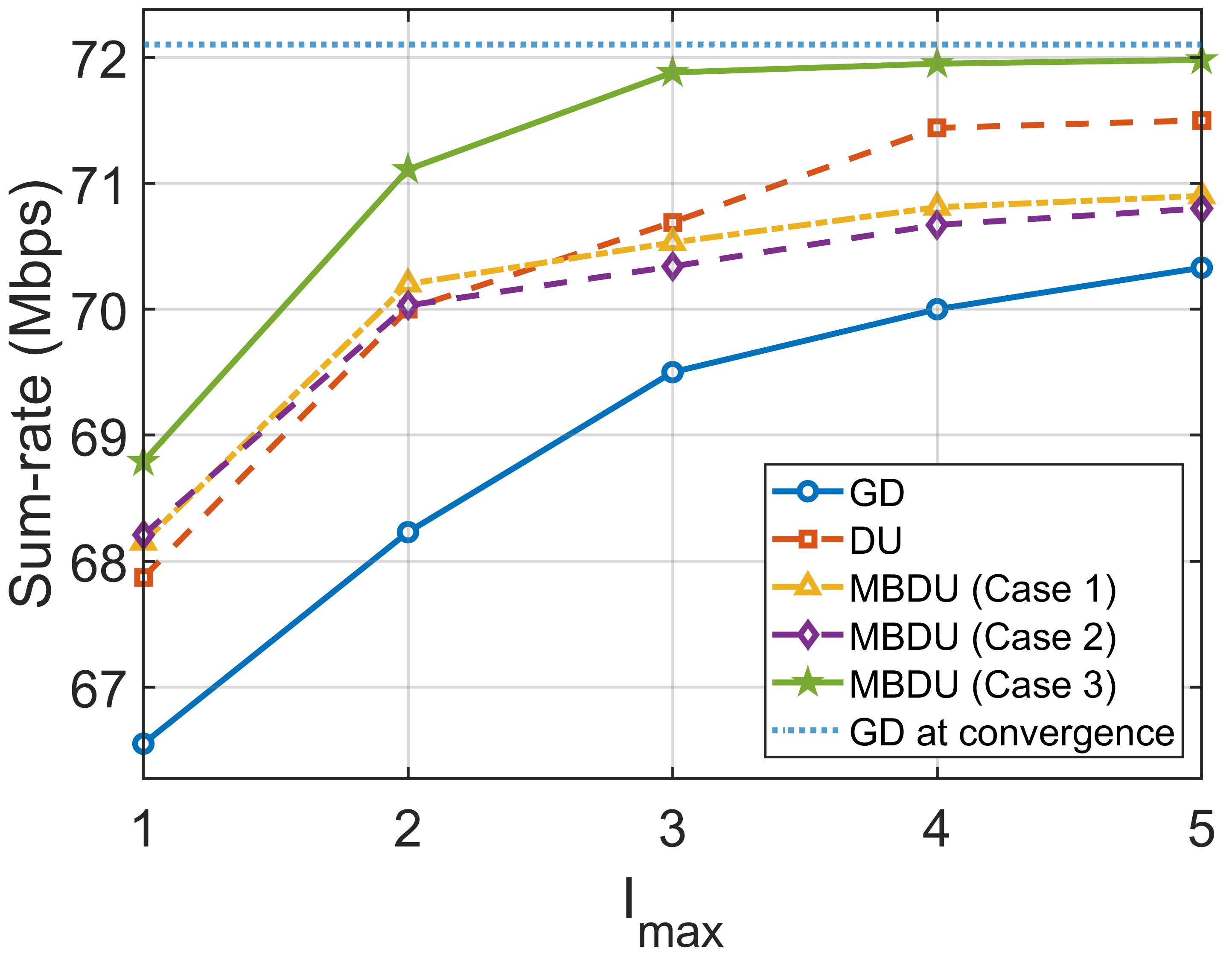}
        \phantomsubcaption\label{fig:I_vs_sumrate_T6}
        \vspace{-1pt}
        {\centering\small (b)\par}
    \end{subfigure}
    \caption{Sum rate versus the number of iterations: (a) Varying the number of inner iterations for $I_{\max}=1$, (b) Varying the number of outer iterations for $T=6$.}
    \label{fig:GD_DU_MBDU_compare}
\end{figure}

We consider a two-carrier multi-band system with a low band and a high band centered at \(f_L=3.5~\mathrm{GHz}\) and \(f_H=17.5~\mathrm{GHz}\), respectively. The subcarrier bandwidths are \(B_L=120~\mathrm{kHz}\) and \(B_H=480~\mathrm{kHz}\), and each band contains \(M_L=4\) and \(M_H=4\) subcarriers unless stated otherwise. The BS total downlink transmit power is fixed to \(P_T=10~\mathrm{W}\). Since the noise power scales linearly with the occupied bandwidth, the high-band noise variance is set larger than the low-band noise variance by the same factor as the bandwidth ratio. The distance between the BS antenna array and the last layer of the SIM is set to \(5\lambda_L\), where \(\lambda_L\) denotes the wavelength at \(f_L\). Unless stated otherwise, we consider \(K=2\) single-antenna users and a BS equipped with \(N_t=2\) active antennas. The SIM has \(L=2\) transmissive layers, each consisting of a \(7\times 7\) planar array of meta-atoms. {For the deep-unfolding-based methods, training is performed offline using independently generated channel realizations with the above transmit-power and noise settings. The trainable step sizes are initialized to \(0.1\), the momentum and extrapolation parameters are initialized to \(0.2\) and \(0.5\), respectively, and the SIM phases are initialized randomly over \([0,2\pi)\). The Adam optimizer is used with a learning rate of \(1\times10^{-3}\), and each model is trained for 40 epochs. The classical GD baseline uses a fixed step size of \(0.1\). For Fig.~\ref{fig:GD_DU_MBDU_compare}, separate DU and MBDU models are trained for each operating point corresponding to the considered values of \(T\) and \(I_{\max}\).}

We compare three SIM phase update methods within the same AO framework: (i) classical projected gradient ascent (GD) with a fixed step size, (ii) basic deep unfolding (DU) obtained by unrolling $T$ projected-gradient updates with trainable step sizes, and (iii) the proposed multi-band deep unfolding (MBDU), which augments the unfolded updates with a momentum state and separate trainable step sizes for the low-band and high-band gradient contributions. For all three methods, the phase update is performed over $I_{\max}$ outer iterations, and performance is evaluated on unseen channels.

Fig.~\ref{fig:GD_DU_MBDU_compare} shows the sum-rate performance under variations of the inner and outer iterations. In Fig.~\ref{fig:GD_DU_MBDU_compare}\subref{fig:T_vs_sumrate_I1}, we fix the number of outer iterations to \(I_{\max}=1\) and vary the inner iterations \(T\). In Fig.~\ref{fig:GD_DU_MBDU_compare}\subref{fig:I_vs_sumrate_T6}, we instead fix the inner iterations to \(T=6\) and vary the maximum number of outer iterations \(I_{\max}\). In both cases, the classical GD baseline improves slowly, indicating that many AO iterations are needed to reach a good operating point. By learning per-stage step sizes, DU achieves faster convergence than GD. {To further examine the MBDU components, we also include three cases. In Case 1, the momentum and extrapolation parameters are fixed, while the band-aware step sizes are learnable. In Case 2, the step sizes are fixed, while the momentum and extrapolation parameters are learnable. In Case 3, all MBDU parameters are learnable. The results show that separately optimizing the band-aware step-size and momentum-related update parameters yields lower sum-rate performance than the jointly learned design. Case 3, where all MBDU update parameters are learnable, achieves the fastest convergence and highest sum rate.}

\begin{figure}[t]
    \centering
    \includegraphics[width=0.9\linewidth]{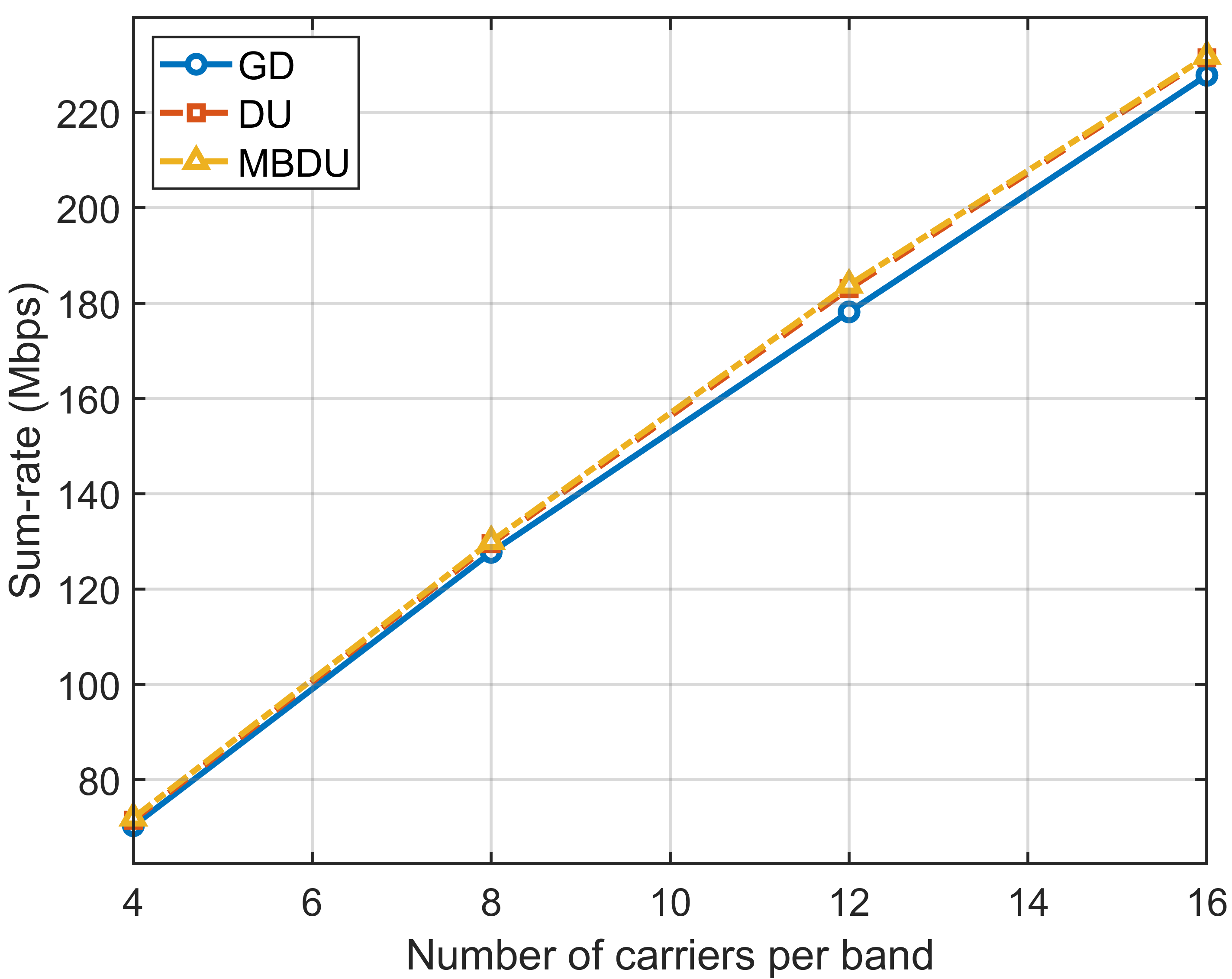}
    \caption{Sum-rate versus the number of subcarriers per band.}
    \label{fig:ML_vs_sumrate}
\end{figure}

Fig.~\ref{fig:ML_vs_sumrate} evaluates robustness with respect to the number of subcarriers per band for $T = 6$ and $I_{\max} = 5$. {The unfolded models are trained using \(M_L=M_H=4\) and then tested on other subcarrier configurations without retraining.} The GD baseline remains consistently below the unfolded approaches across the entire range, indicating that the unfolded methods generalize well on unseen data.
%--------------------------------------------------------------------------
\vspace{-5pt}
\section{Conclusion}
We have investigated SIM-assisted multi-band multiuser downlink transmission under a physically consistent impedance-domain SIM model, where carrier aggregation introduces heterogeneous subcarrier conditions across bands, while the SIM
phase shift remains shared. This coupling makes conventional alternating-optimization solvers costly because the SIM phase update often requires many iterations. To address this limitation, we have proposed MBDU-Net, a physically-consistent multi-band deep-unfolding network for SIM phase optimization. MBDU-Net unrolls a projected-gradient phase update into a fixed depth architecture that retains the analytic gradient structure of the SIM channel model, while learning lightweight parameters that govern the update dynamics, including band-aware scaling and momentum. Numerical results show that the proposed unfolding provides suboptimal performance with substantially fewer outer iterations than classical projected-gradient methods, and maintains strong performance when the number of subcarriers differs from the training configuration.

\bibliographystyle{IEEEtran}
\bibliography{paper_refs.bib}

\end{document}